%% file: main.tex
\documentclass[review]{cas-dc}
\usepackage{graphicx}
\usepackage{rotating}
\usepackage{amsmath,amsfonts}
\usepackage[numbers]{natbib}

\newcommand\revise[1]{\textcolor{black}{#1}}
\newcommand{\Z}{\mathcal{Z}}
\newcommand{\T}{\mathcal{T}}
\newcommand{\J}{L_{\text{control}}}
\newcommand{\R}{\mathbb{R}}
\newcommand{\tr}{{{\mathsf T}}}
\usepackage{algorithm}
\usepackage{algpseudocode}
\hypersetup{citecolor=blue, urlcolor=blue}
\usepackage{float}

\begin{document}
\let\WriteBookmarks\relax
\def\floatpagepagefraction{1}
\def\textpagefraction{.001}

\shorttitle{Ventilation and Temperature Control for Energy-efficient and Healthy Buildings: A Differentiable PDE
Approach}

\shortauthors{Yuexin Bian et~al.}

\title [mode = title]{Ventilation and Temperature Control for Energy-efficient and Healthy Buildings: A Differentiable PDE
Approach}                      

\tnotetext[1]{This work is funded by the Hellman Fellowship and University of California San Diego.}


\author[1]{Yuexin Bian}
\affiliation[1]{{Department of Electrical and Computer Engineering, University of California San Diego},
    addressline={}, 
    city={La Jolla},
    postcode={92037}, 
    country={USA}}
\credit{Investigation, Methodology, Software, Writing – original draft, Writing - review \& editing}

\author[2]{Xiaohan Fu}
\affiliation[2]{{Department of Computer Science and Engineering, University of California San Diego},
    addressline={}, 
    city={La Jolla},
    postcode={92037}, 
    country={USA}}
\credit{Investigation, Writing - review \& editing}
\author[2]{Rajesh K. Gupta}
\credit{Supervision, Writing - review \& editing}
\author[1]{Yuanyuan Shi}[
orcid=0000-0002-6182-7664
]
\credit{Supervision, Methodology, Writing - review \& editing}
\cormark[1]

\cortext[cor1]{Corresponding author}



\begin{abstract}
In response to the COVID-19 pandemic, there has been a notable shift in literature towards enhancing indoor air quality and public health via Heating, Ventilation, and Air Conditioning (HVAC) control. However, many of these studies simplify indoor dynamics using ordinary differential equations (ODEs), neglecting the complex airflow dynamics and the resulted \emph{spatial-temporal} distribution of aerosol particles, gas constituents and viral pathogen, 
which is crucial for effective ventilation control design. We present an innovative partial differential equation (PDE)-based learning and control framework for building HVAC control. 
The goal is to determine the optimal airflow supply rate and supply air temperature to minimize the energy consumption while maintaining a comfortable and healthy indoor environment.
In the proposed framework, the dynamics of airflow, thermal dynamics, and air quality (measured by CO$_2$ concentration) are modeled using PDEs. 
We formulate both the system learning and optimal HVAC control as PDE-constrained optimization, and we propose a gradient descent approach based on the adjoint method to effectively learn the unknown PDE model parameters and optimize the building control actions. We demonstrate that the proposed approach can accurately learn the building model on both synthetic and real-world datasets. Furthermore, the proposed approach can significantly reduce energy consumption
while ensuring occupants' comfort and safety constraints compared to existing control methods such as maximum airflow policy, learning-based control with reinforcement learning, and control with ODE models. 
\end{abstract}



\begin{keywords}
\sep Energy-efficient buildings
\sep Data-driven control
\sep Heating, ventilation, and air conditioning (HVAC) system
\sep Indoor air quality
\sep Control of partial differential equations
\end{keywords}

\maketitle

\section{Introduction}\label{sec:1}
\input{introduction}

\section{System model}\label{sec:2}
\input{formulation}

\section{Solution approach}\label{sec:4}
\input{method}

\section{Numerical experiments}\label{sec:5}
\input{experiment}

\section{Conclusion and future work}\label{sec:6}
In this paper, we introduce a novel framework for building learning and control, focusing on ventilation and thermal management to enhance energy efficiency. We validate the performance of the proposed framework in system model learning via two case studies: a synthetic study focusing on the joint learning of temperature and CO$_2$ fields, and an application to a real-world dataset for CO$_2$ field learning. For building control, we demonstrate that the proposed framework can optimize the control actions and significantly reduce the energy cost while maintaining a comfort and healthy indoor environment. When compared to existing traditional methods, an optimization-based method with ODE models and reinforcement learning, our approach can significantly reduce the energy consumption while guarantees all the safety-critical air quality and control constraints. 
Promising future research directions involve validating and improving the proposed PDE models through accurate estimation of airflow fields within indoor environments. Additionally, incorporating uncertainty modeling into the PDE framework for HVAC control presents an opportunity to enhance the efficiency and reliability of building HVAC system management.

\appendix
\input{appendix}

\printcredits

\bibliographystyle{elsarticle-num}
\bibliography{reference}

\end{document}

%% file: introduction.tex
\subsection{Background and literature review}
Buildings are a significant contributor to energy consumption, accounting for over 40\% of the global energy use~\cite{harputlugil2021interaction}. HVAC systems 
account for up to 50\% of a building's energy usage~\cite{che2019energy}, in both commercial and residential buildings. Consequently, there is a growing demand for optimizing energy consumption and lowering carbon footprints of building HVAC systems. 
When optimizing building energy consumption, it is important to account for the comfort and safety requirements of human occupants. According to Mannan et al.~\cite{mannan2021indoor}, people spend approximately 90\% of their lifetime indoors. The quality of indoor environments becomes critical for human health and productivity in buildings, however, maintaining comfortable temperature and healthy air quality may lead to increased energy consumption~\cite{boodi2018intelligent,xia2023reca}. 

In the post-COVID era, HVAC energy management has become even more challenging.
According to~\cite{hosseinloo2023data,dong2022optimal}, high airflow rates can reduce the exposure of occupants to viral pathogens in indoor environments thus reducing the infection risks. Yet, this necessary increase of airflow rate can lead to higher energy consumption.
In practice, many HVAC systems have been operating at maximum airflow rates in response to COVID-19. For instance, starting from the spring 2020, the Facilities Management (FM) at UC San Diego has implemented a policy of maximum fresh-air intake with minimal or no recirculation during office hours~\cite{UCSD_facility}, which results in the building's energy consumption being 2-2.5 times higher than the nominal energy costs. 
The European REHVA~\cite{REHAV} also suggested that the HVAC system should operate at a high air supply rate and exhaust ventilation rate, while adjusting the setpoint of the CO$_2$ concentration to 400 ppm. 
While beneficial for air quality and reducing infection risks, these HVAC control policies are unsustainable for various reasons, including skyrocketing energy consumption and strain on the mechanical systems. 
Therefore, it is imperative to develop an integrated control framework that simultaneously ensures a comfortable and healthy indoor environment while minimizing the energy consumption. 

In this study, we utilize CO$_2$ concentrations as an indicator of indoor air quality.
As noted by Schibuola et al.~\cite{schibuola2021high}, direct information on indoor viral distribution is typically unavailable. Instead, CO$_2$ concentrations are used as an practical measure to infer air changes, which can help estimate viral concentration and associated infection risk~\cite{schibuola2021high}. Moreover, Shinohara at al.~\cite{shinohara2022air} has shown that with air conditioning(AC) off, CO$_2$ and aerosol particles spread proportionally at the same rate from the source, whereas with the AC on, the spread rate of particles was about half that of CO$_2$. This observation underpins the use of CO$_2$ constraints in various studies~\cite{schibuola2021high, li2022tube, li2020multi, zhang2021novel, boodi2018intelligent} as a means to maintain a healthy indoor environment.

Existing building HVAC control methods fall into three categories: rule-based control, optimization-based control, and learning-based control. Rule-based control has been widely deployed in most real-world building systems~\cite{gwerder2010potential,bian2023bear}. For example, Bian et al.~\cite{bian2023bear} showed that a majority of UC San Diego campus buildings are operated with rule-based control. Jiang et al.~\cite{jiang2023pandemic} set the airflow rate using the Wells-Riley model~\cite{riley1978airborne}, \revise{where the desired airflow rate depends on the number of occupants, infection risk, whether people are wearing masks and other factors. }
However, rule-based control requires intensive human work and domain knowledge to generate rules. Additionally, applying the rule-based control for multi-objective building control problems is challenging~\cite{chen2023adaptive}, to trade-off different objectives while satisfying all constraints. 
Optimization-based methods~\cite{mayer2016branch, yang2020hvac, chen2018optimal, li2022tube} formulate the HVAC control as an optimization problem, where the objective function can be customized (energy consumption, operation cost, among others), subject to the building dynamics model and state/action constraints. See~\cite{drgovna2020all} for a recent review about optimization-based approaches for building control.

Recent advancements in machine learning have opened new avenues for energy optimization and autonomous operation of building systems. At the same time, more data is becoming available due to the widespread deployment of smart sensors. As a result, learning-based control techniques~\cite{maddalena2020data}, especially reinforcement learning (RL), have attracted surging attention for building control. Researchers~\cite{zhang2023bear,wei2017deep,hosseinloo2023data,yu2020multi,wang2024energy,du2021intelligent} utilized RL to develop optimal control for energy cost minimization, thermal comfort control, viral pathogen prevention, and more. However, RL methods require extensive training data and lack hard constraint satisfaction in deployment. For example, it is noted that an RL agent might require up to 5 million interaction steps, equivalent to 47.5 years of simulated data, to match the performance of a traditional feedback controller in an HVAC system~\cite{zhang2018practical}.

In our work, we take an optimization-based control approach. The building HVAC dynamics are modeled as partial differential equations (PDEs) whose parameters will be determined from data, and the optimal control actions are derived from solving the PDE-constrained optimization. This approach significantly reduces the need for extensive historical data, while guaranteeing satisfaction of hard constraints on health/safety states and control actions.

\subsection{Related work in HVAC control and air quality}
HVAC control is designed to ensure occupant comfort and energy efficiency. Yet, the majority of previous works have focused on studying the thermal control of buildings. 
Boodi et al.~\cite{boodi2018intelligent} 
noted that while 84\% of the literature considers thermal comfort and energy efficiency, only 5\% of studies addresses the indoor air quality. There remains a notable gap in the advancement of airflow control strategies, particularly in relation to airflow control and indoor air quality. 

In the wake of the COVID-19 pandemic, the focus has shifted towards more in-depth investigations into improving indoor air quality and reducing infection risks through HVAC control. For instance, Li et al.~\cite{li2022tube} modeled CO$_2$ dynamics using an ordinary differential equation (ODE) and solved the optimization problem to minimize energy usage while ensuring good indoor air quality. Zhang et al.~\cite{zhang2021novel} employed an ODE model to represent CO$_2$ dynamics, solving the HVAC control problem with a genetic algorithm.  
Li et al.~\cite{li2020multi} modeled the CO$_2$ level with a mass balance equation and tackled the optimal control problem with a distributed optimization approach. However, 
ODE models for CO$_2$ (and aerosol particles) concentrations compromise accuracy in modeling spatial-temporal airflow dynamics. This limitation can lead to inferior air quality and increased infection risks in certain areas, making these models inadequate for effective ventilation and airflow design~\cite{hosseinloo2023data, he2016zoned}.

To model the airflow dynamics, computational fluid dynamics
(CFD) simulations were adopted. For instance, Lau et al.~\cite{lau2022predicting} utilized advection–diffusion–reaction equations to predict the spatial-temporal infection risk. However, their focus was solely on assessing infection risk without considering how ventilation control could mitigate these risks. Narayanan et al.~\cite{narayanan2021airborne} solved Navier-Stokes equations coupled with a transport equation for spatiotemporal pathogen concentration in a music classroom. Although their research offers insights into the effects of using portable air purifiers, it did not optimize room control strategies. Additionally, Jin et al.~\cite{jin2015sensing} utilized a convection PDE to model the CO$_2$ concentration, yet their analysis was limited to scenarios with a constant airflow rate, overlooking the influence of varying ventilation rates on airflow dynamics. 
Koga et al.~\cite{koga2019learning} utilized the Navier-Stokes and convection-diffusion equations to model spatiotemporal airflow and temperature. However, their study was solely focused on identifying model parameters.
Hosseinloo et al.~\cite{hosseinloo2023data} modeled pathogen concentration using convection-diffusion equations and optimized the velocity field through RL. However, directly optimizing the entire velocity field is impractical, as control is limited to the boundary airflow velocity at the supply air vents within a building.
He et al.~\cite{he2016zoned} applied Navier-Stokes and convection-diffusion equations to model spatiotemporal airflow and temperature, addressing indoor temperature control through PDE-constrained optimization. While their approach utilizes Finite Element Method (FEM) discretization and Differential Algebraic Equations (DAE) for optimal control, it is computationally demanding. Such complexity renders it less practical for building management systems that require comprehensive planning over extended periods. Moreover, their approach does not take air quality into consideration.



\begin{figure*}[t]
    \centering
    \includegraphics[width=2.0\columnwidth]{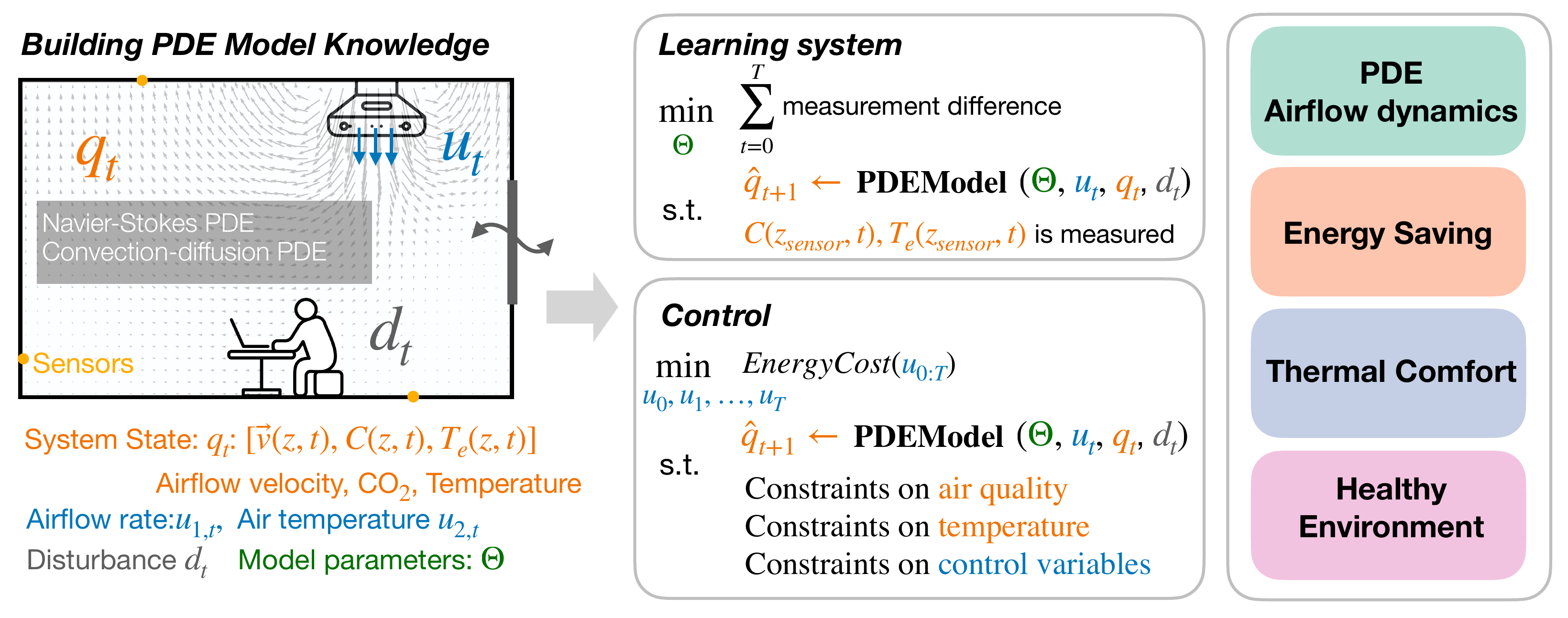}
    \caption{Schematic of the proposed PDE-based learning and control framework for healthy and energy-efficient buildings. \revise{Left figure: we model the airflow, CO$_2$ and temperature dynamics with PDEs, including Navier-Stokes and convection-diffusion PDEs, with the unknown model parameters denoted by $\Theta$. The HVAC control and external disturbances are represented as $u_t$ and $d_t$, respectively. Right figure: we present the formulation of the learning system and control problem as PDE-constrained optimization problems. The goal is to learn unknown system parameters and determine optimal control actions. The results show that the proposed framework is able to model spatiotemporal dynamics with PDEs, leading to improved energy efficiency, ensured thermal comfort, and a healthy indoor environment. } 
    }
    \label{fig:frame}
\end{figure*}

\subsection{Contributions and innovations}
To the best of our knowledge, this is the first PDE-based learning and control 
framework for building control that simultaneously optimizes energy consumption, and guarantees air quality and thermal constraints. 
Our proposed framework is in Figure~\ref{fig:frame}, 
where the system state includes airflow velocity, CO$_2$ concentration~\footnote{We use CO$_2$ concentrations as an indicator of air quality, and it allows for the inclusion of other aerosol particles, e.g., PM 2.5, PM 10 and airborne pathogens~\cite{zhang2020identifying}.}, and temperature. The airflow velocity field is modeled by the Navier-Stokes equations, and the dynamics of CO$_2$ concentration and temperature are modeled with convection-diffusion PDEs.
In the system learning task, the goal is to estimate the \emph{unknown} building parameters that govern the fluid dynamics. In the control task, the goal is to minimize the energy consumption while ensuring thermal comfort and air quality, via optimizing the supply airflow rate and supply air temperature setpoints. 

We propose a gradient descent approach for solving the PDE-constrained optimization for both building model learning and optimal control tasks. 
Our approach achieves a significant reduction in energy consumption, compared to existing control methods such as maximum
airflow policy, learning-based control with RL, and optimization-based control
with ODE models. 
Compared to the maximum airflow policy, our method achieves a 52.6\% reduction in energy consumption. Additionally, we see energy savings of 36.4\% and 10.3\% compared to RL and control with ODE models, respectively. While RL and control with ODE models occasionally violate the safety constraints, our approach successfully maintains comfortable and healthy environmental standards at all time.


We organize the remainder of the paper as follows. Section~\ref{sec:2} and \ref{sec:3} present the building PDE models and problem formulation. Section~\ref{sec:4} provides the solution method for solving the PDE-constrained building model learning and optimal control. Section~\ref{sec:5} presents case studies, detailing the application of our approach to both learning tasks using synthetic and real-world data, and control tasks. Section~\ref{sec:6} concludes the paper and outlines future research directions. 

%% file: formulation.tex
\begin{table}[h]
    \centering
    \caption{Notation used for PDE models.}
    \begin{tabular}{ll}\hline
   Notation     &  Description \\\hline
    $\Z$        &  domain of PDEs \\
    $z = (x,y)$ &  spatial coordinate \\
    $\partial \Z$ & boundary of the field \\
    $\Z_{\text{supply}}$ & air supply vent positions \\
    $\Z_{\text{return}}$ & air return vent positions \\
    $\Z_{\text{outside}}$ & wall adjacent to the exterior\\ 
    $\T=[0, \overline{T}], \, t$  & the time set and time index \\ 
    $\vec{v}(z,t)$ & airflow velocity field \\
    $p(z,t)$ & pressure field  \\
    $T_e(z,t), d_{T_e}(z,t)$ & temperature field and heat source \\
    $C(z,t), d_C(z,t)$ & CO$_2$ field and CO$_2$ source \\ 
    $g(z,t)$ & the number of people in a room\\ 
    $q_t$ & system state $q_t = [\vec{v}(z,t), C(z,t), T_e(z,t)]$ \\
    $T_{\text{ambient}}(t)$ & ambient temperature \\
    $\rho$ & fluid density \\
    $C_{\text{fresh}}$ & CO$_2$ level for fresh air in~\eqref{eq:c1}\\
    \hline
    Model parameters & \\ 
    $\nu$ & kinematic viscosity \\  
    $k_{T_e}$ & diffusion coefficient of temperature\\  
    $k_C$ & diffusion coefficient of CO$_2$   \\ 
    $\alpha$ & the re-circulation rate in~\eqref{eq:c1}\\
    $\alpha_{T_e}$ & heat source coefficient in~\eqref{eq:occupant} \\
    $\alpha_C$ & CO$_2$ source coefficient in~\eqref{eq:occupant} \\
    \hline
    Control variables & \\
    $u_{1,t} \in \R$ & supply airflow rate \\
    $u_{2,t} \in \R$ & supply air temperature \\
    \hline
    \end{tabular}
    \label{tab:notation}
\end{table}

In this section, we describe the building dynamic models that characterize the airflow velocity, temperature field, and CO$_2$ concentration field, which are modeled by PDEs. 
The notations are summarized in Table~\ref{tab:notation}.
\begin{figure*}
    \centering
    \includegraphics[width=1.9\columnwidth]{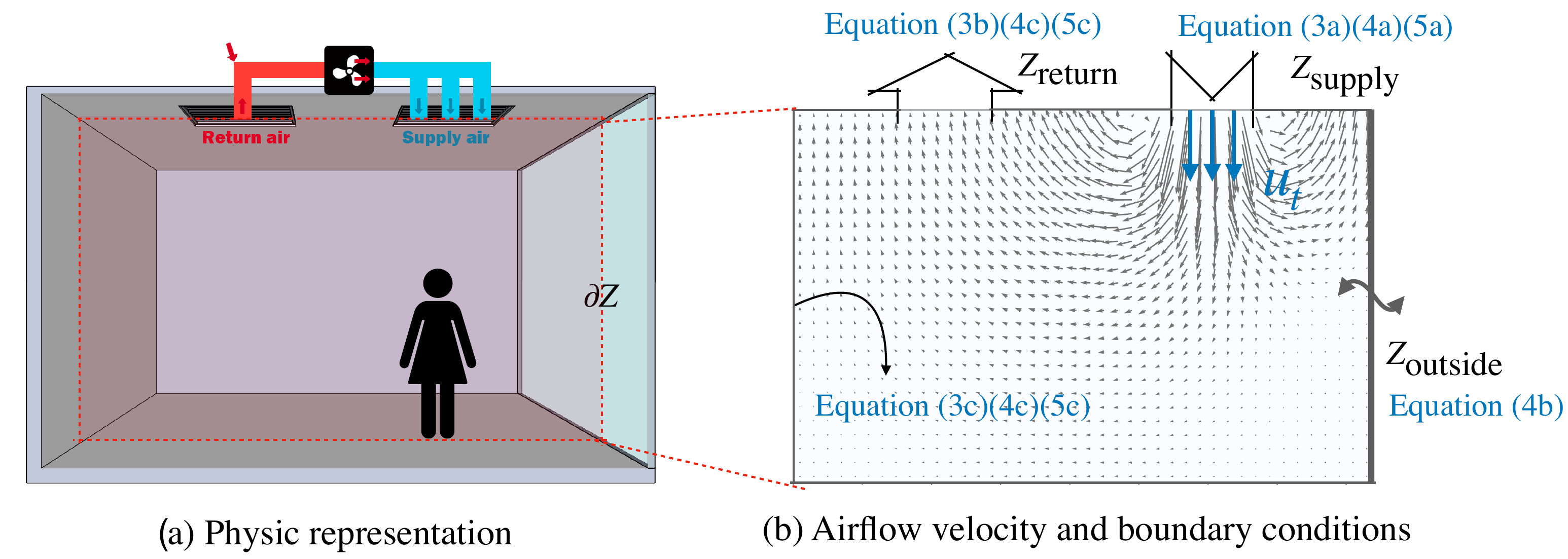}
    \caption{(a) The physic representation of the simulation testbed: a typical room with a ventilation system including air supply and air return vents on the ceiling. \revise{All the walls are insulated, and the right side features a glass window wall.}
    (b) We define a 2D region and model it using 2D PDEs. The figure visualizes the airflow velocity, governed by the Navier-Stokes equations. All boundary conditions are highlighted in blue text.
    }
    \label{fig:boundary}
\end{figure*}

\subsection{Models for airflow, temperature and CO$_2$}
\label{sec:pde}
Let us consider, a typical office environment where the temperature and CO$_2$ levels dynamically evolve due to a set of contributing factors including the physical layout of the space, HVAC control actions (supply airflow rate and temperature), human occupancy, and external weather conditions. The physical representation of considered room is illustrated in Figure~\ref{fig:boundary}. This room bears close resemblance to other typical indoor spaces, with a ventilation system including air supply and air return vents on the ceiling. 

We employ PDEs to model the spatiotemporal dynamics of the airflow velocity field, and the resulted temperature and CO$_2$ fields.   
We denote the domain of PDE by $\Z \subset \R^2$ (or $\R^3$), which represents a confined region, e.g. a meeting room. 
The time domain is $\mathcal{T} = [0, \overline{T}] \subset \R^+$. Spatial coordinate and time are denoted by $z \in \Z$ and $t \in \T$.
We denote the airflow velocity field $\vec{v}(z,t): \Z \times \T \rightarrow \R^2$, the temperature field $T_e(z,t):\Z \times \T \rightarrow \R$ and the CO$_2$ field $C(z,t): \Z \times \T \rightarrow \R$. 

The airflow velocity field is modeled by the Navier–Stokes equations,
\begin{subequations}\label{eq:ns}
    \begin{align}
     & \nabla \cdot \vec{v} = 0, & \text{Continuity Equation}\label{eq:v1} \\
     & \frac{\partial \vec{v}}{\partial t} + \vec{v} \cdot \nabla \vec{v} = -\frac{1}{\rho}\nabla p + \nu \nabla^2 \vec{v}  + \vec{g}, & \text{Momentum Equation}\label{eq:v2} 
    \end{align}
\end{subequations}
where $\vec{v}$ is the velocity vector, $\nu$ is the kinematic viscosity of the airflow, $\rho$ is the fluid density, $p$ is the pressure field and $\vec{g} = [0, -9.8]^\tr$ is the gravitational force.

The thermal and CO$_2$ dynamics are modeled by the convection-diffusion equations, 
\begin{subequations}\label{eq:tc}
\begin{align}
  &  \frac{\partial T_e}{\partial t} + \vec{v} \cdot \nabla T_e = k_{T_e} \nabla^2 T_e + d_{T_e}(z,t), & \text{Temperature} \label{eq:t0} \\
   &  \frac{\partial C}{\partial t} + \vec{v} \cdot \nabla C = k_C\nabla^2 C + d_C(z,t), & \text{CO$_2$}\label{eq:c0}
\end{align}
\end{subequations}
where $k_{T_e}, k_C$ are the diffusion coefficients for the temperature and CO$_2$ fields, and $d_{T_e}(z,t), d_C(z,t)$ are the heating source and CO$_2$ source, respectively. We employ straightforward proportional models to represent both the heat and CO$_2$ sources in a room~\cite{hosseinloo2023data, xiao2023building, bian2022demand},
\begin{equation}\label{eq:occupant}
    d_{T_e}(z,t) = \alpha_{T_e} g(z,t)\,, \quad d_C(z,t) = \alpha_C g(z,t)\,,
\end{equation}
where $g(z,t)$ denotes the number of people at position $z$ at time step $t$ and $\alpha_{T_e}, \alpha_C$ are the respective coefficients for heat and CO$_2$ source per occupant. Our framework can also take more sophisticated models of occupants' effects, such as polynomial models~\cite{zhang2023bear}.

\subsection{Boundary conditions}
In this work, we consider HVAC control as boundary control~\cite{hosseinloo2023data,he2016zoned}, with the specific boundary conditions detailed in Figure~\ref{fig:boundary} (b).  
The boundary of the field is denoted by $\partial \mathcal{Z}$, which represents the walls, the ceiling and the ground of the room. Within this boundary, the positions of the air supply vent and air return vent are specified as $\Z_{\text{supply}}$ and $\Z_{\text{return}}$. We denote the control variable $u_t \in \mathbb{R}^2$, where $u_{1,t} \in \mathbb{R}$ represents the airflow rate at the supply vent (pointing in a downward direction), and $u_{2,t} \in \mathbb{R}$ represents the supply air temperature at the supply vent. 
The boundary conditions for the airflow velocity field, temperature field, and CO$_2$ field are defined below. 
\subsubsection{Boundary conditions for airflow velocity field}
Boundary conditions for airflow velocity field are defined as,
\begin{subequations}\label{eq:boundary_v}
    \begin{align}
     & \vec{v}(z,t) = -u_{1,t}  \cdot e_y, \forall z \in \Z_{\text{supply}}, \label{eq:u1}\\
      &  \vec{n} \cdot \nabla \vec{v} = 0, \forall z \in \Z_{\text{return}},      \label{eq:u2_2}\\
     & \vec{v} = 0, \forall z \in \partial \Z \setminus (\Z_{\text{supply}} \cup \Z_{\text{return}}).  \label{eq:u2} 
    \end{align}
\end{subequations}
Constraint \eqref{eq:u1} specifies that the airflow velocity at the supply vent $\Z_{\text{supply}}$ is controlled by $u_{1,t}$ (m/s), where $e_y = [0, 1]^\top$ is the unit vector in the $y$-upward direction. Constraint~\eqref{eq:u2_2} sets the Neumann boundary conditions at the return vent and constraint~\eqref{eq:u2} applies Dirichlet conditions to all other boundaries by setting the airflow velocity as zero~\cite{he2016zoned}.

\subsubsection{Boundary conditions for temperature field}
Boundary conditions for the temperature field are defined as,
\begin{subequations}\label{eq:boundary_t}
    \begin{align}
     & T_e(z, t) = u_{2,t}, \forall z \in \Z_{\text{supply}},\label{eq:t1}\\
     & T_e(z, t) = T_{\text{ambient}}(t), \forall z \in \Z_{\text{outside}}, \label{eq:t1_2}\\
     & \vec{n} \cdot \nabla T_e = 0, \forall z \in \partial \Z  \setminus (\Z_{\text{supply}} \cup \Z_{\text{outside}}). \label{eq:t2}
    \end{align}
\end{subequations}
Constraint~\eqref{eq:t1} states that the temperature at the supply air vent is controlled as $u_{2,t}$ ($^\circ$C), and constraint~\eqref{eq:t1_2} represents the temperature of the right window wall $\Z_{\text{outside}}$ is influenced by the ambient temperature $T_{\text{ambient}}(t)$. Constraint~\eqref{eq:t2} sets the Neumann boundary conditions for all the solid walls as insulated surfaces~\cite{farahmand2017deep}.
\subsubsection{Boundary conditions for CO$_2$ field}
Boundary conditions for the CO$_2$ field are defined as,
\begin{subequations}\label{eq:boundary_c}
    \begin{align}
    &C(z,t) = \alpha \cdot C_{\text{fresh}}  +  (1-\alpha) \cdot \frac{1}{\Z} \int_{\Z} C(z,t) \text{d} z, \forall z \in \Z_{\text{supply}}, \label{eq:c1}\\
    &\vec{n} \cdot \nabla C = 0, \forall z \in \partial \Z \setminus\Z_{\text{supply}}. \label{eq:c2}
    \end{align}
\end{subequations}
Constraint \eqref{eq:c1} dictates that the CO$_2$ concentration at the supply air vent is a \emph{mixture} of the CO$_2$ concentration of fresh air $C_{\text{fresh}}$ and the CO$_2$ concentration of recirculated air within the building $\frac{1}{\Z} \int_{\Z} C(z,t) \text{d} z$. $\alpha$ represents the re-circulation rate that can vary among different buildings. In this work, the fresh air CO$_2$ concentration $C_{\text{fresh}}$ is set as 400 ppm~\cite{li2022tube} and $\alpha$ is a parameter to be identified from data. 
\revise{Constraint~\eqref{eq:c2} establishes the Neumann boundary conditions for all the boundaries except the supply vent~\cite{hosseinloo2023data}.}

\section{Problem formulation}\label{sec:3}
\revise{Given the PDE model knowledge described in Section~\ref{sec:2}, we formulate the system learning and control problem as PDE-constrained optimization problems. The goal is to learn the unknown parameters in the PDE system and develop control algorithms to optimize the energy efficiency, thermal comfort and indoor air quality. }

\label{sec:formulation}
\subsection{Learning the system model}
A fundamental challenge in controlling building systems is that the relationship between the airflow, temperature, and CO$_2$ concentration and the HVAC control actions is governed by a set of nonlinear PDEs as described in Section \ref{sec:2}, whose parameters depend on detailed building characteristics
that are difficult to measure in practice.
We consider the set of system model parameters to be identified include \revise{$\Theta = \{\nu, k_{T_e}, k_C, \alpha, \alpha_{T_e}, \alpha_C\}$. }
It is important to note that for incompressible flow, the fluid density $\rho$ in \eqref{eq:v2} is considered constant, thus it has no impacts on the velocity, temperature and CO$_2$ concentration values. Pressure computation adjusts the velocity field to meet the continuity equation, ensuring incompressibility. 
Given historical temperature and CO$_2$ records obtained from sensors placed at specific locations, our goal is to learn the unknown system parameters $\Theta$, which minimize the difference between actual historical CO$_2$/Temperature records and predicted CO$_2$/Temperature values based on estimated parameters,
\begin{subequations}\label{eq:p_sysid}
\begin{align}
     \min_{\Theta} & \quad \sum_t (\| T_e(z_{sensor},t)  - \widehat{T}_e(z_{sensor},t) \|^2 \label{eq:l_cost1}\\
      &   + \| C(z_{sensor},t)  - \widehat{C}(z_{sensor},t) \|^2)
     \label{eq:l_cost2}\\
     \text{s.t.}\ \  & {\color{black} \Theta = \{\nu, k_{T_e}, k_C, \alpha, \alpha_{T_e}, \alpha_C\}}, \\ 
     & \nabla \cdot \widehat{\vec{v}} = 0, \\
    & \frac{\partial \widehat{\vec{v}}}{\partial t} + \widehat{\vec{v}} \cdot \nabla \widehat{\vec{v}} = -\frac{1}{{\rho}}\nabla \hat{p} + {\color{black} \nu} \nabla^2 \widehat{\vec{v}}  + \vec{g}, \\
    & \frac{\partial \widehat{T_e}}{\partial t} + \widehat{\vec{v}} \cdot \nabla \widehat{T_e} = {\color{black}{k}_{T_e}} \nabla^2 \widehat{T_e} + {\color{black}{\alpha_{T_e}}} g(z,t), \\
    & \frac{\partial \widehat{C}}{\partial t} + \widehat{\vec{v}} \cdot \nabla \widehat{C} = {\color{black}{k}_C}  \nabla^2 \widehat{C} +  {\color{black}{\alpha_C}} g(z,t), \\
    & \small{\widehat{C}(z, t) = {\color{black} \alpha} \cdot C_{\text{fresh}}  +  (1-{\color{black} \alpha}) \cdot \frac{1}{\Z} \int_{\Z} \widehat{C}(z,t) \text{d} z\,, \forall z \in \Z_{\text{supply}}}, \\
    &~\eqref{eq:boundary_v}\eqref{eq:boundary_t}\eqref{eq:c2} \tag{\text{other boundary conditions}} 
\end{align} 
\end{subequations}
where $z_{sensor} \subset \Z$ represents the set of sensor positions. The predicted temperature and CO$_2$ $\widehat{T}_e(z_{sensor},t), \widehat{C}(z_{sensor},t)$ are driven by the PDEs with the estimated parameters $\Theta = \{\nu, k_{T_e}, k_C, \alpha, \alpha_{T_e}, \alpha_C\}$, and $T_e(z_{sensor},t)$, $C(z_{sensor},t)$ are the ground-truth temperature and CO$_2$ measurements recorded by sensors.

\subsection{Control problem formulation}
Now we are able to formulate the HVAC control problem. We note that the airflow dynamics~\eqref{eq:ns}, temperature and CO$_2$ dynamics~\eqref{eq:tc} and equation~\eqref{eq:c1} are driven under the estimated parameters obtained by solving the system learning problem~\eqref{eq:p_sysid}.
\begin{subequations}\label{eq:p_control}
\begin{align}
     \min_{u_t} \quad & \int_{t=0}^{\overline{T}} l_{\text{control}}(t)\,  \text{d}t
     \label{eq:c_cost}\\
\text{s.t.} \quad & ~\eqref{eq:ns} \tag{\text{Airflow dynamics}} \\
    & ~\eqref{eq:tc} \tag{\text{Temperature and CO$_2$ dynamics}} \\
    & ~\eqref{eq:boundary_v}~\eqref{eq:boundary_t}~\eqref{eq:boundary_c} \tag{\text{Boundary conditions}} \\
    & \underline{u_1} \leq u_{1,t} \leq \overline{u_1}, \underline{u_2} \leq u_{2,t} \leq \overline{u_2}, \forall t, \label{eq:cons1}\\
    & C(z, t) \leq C_{\text{max}},   \forall z \in \mathcal{Z}, t \in \mathcal{T},\label{eq:cons2} \\
    &  T_{\text{min}} \leq  \frac{1}{\Z} \int_z T_e(z,t)\text{d}z \leq T_{\text{max}},   \forall t \in \mathcal{T}. \label{eq:cons3}
\end{align} 
\end{subequations}
\revise{We define the loss objective as an integral of $l_{\text{control}}(t)$ over time $t$. $l_{\text{control}}(t)$ can include costs such as energy consumption, comfort score that reflects indoor air quality and temperature, operational expenses, among others. }
Constraint~\eqref{eq:cons1} are the control action constraints, where $\overline{u_1}$ and $\underline{u_1}$ represent the maximum and minimum supply airflow rates, and $\overline{u_2}$ and $\underline{u_2}$ denote the maximum and minimum supply airflow temperatures.
Constraints \eqref{eq:cons2} and \eqref{eq:cons3} are designed to ensure that CO$_2$ concentration levels do not exceed healthy limits at \emph{all} locations and that the average room temperature remains within occupants' thermal comfortable range. \revise{Average temperature is a practical measure for occupant comfort since people are less sensitive to minor temperature variations. However, CO$_2$ are strictly controlled everywhere due to higher sensitivity and infection risks associated with poor air quality~\cite{shinohara2022air}. }

To handle the control constraint~\eqref{eq:cons1}, we use a projected gradient method to ensure the constraint. 
To deal with the state constraints~\eqref{eq:cons2}\eqref{eq:cons3}, we utilize log barrier functions for the inequality constraints, 
\begin{equation}\label{eq:control_cost}
\begin{aligned}
   \bar{L}_{\text{control}} &= \int_{t=0}^{\overline{T}} [l_{\text{control}}(t) - \alpha_1 \log \left( C_{\text{max}} - \max_{z}(C(z,t)) \right) \\
    & - \alpha_2 \log \left(T_{\text{max}} - \frac{1}{\Z} \int_z T_e(z,t) \text{d}z\right) \\
    & - \alpha_2 \log \left(\frac{1}{\Z} \int_z T_e(z,t)\text{d}z-T_{\text{min}}\right) ] \, \text{d}t
\end{aligned}
\end{equation}
where $\alpha_1, \alpha_2$ are the weight factors.

\revise{In this study, we define $l_{\text{control}}(t)$ to account for both energy consumption and deviations in control variables. The first term aims to minimize energy usage, while the second term enhances the practical applicability of the control for building management systems. 
\begin{equation}\label{eq:lcontrol}
    l_{\text{control}}(t) = Energy(t) + \alpha_3 \dot{u}_t
\end{equation}
$\alpha_3$ is the weight factor. For the energy consumption, we use the L1 norm of airflow rate $u_{1,t}$ and the difference between the supply air temperature $u_{2,t}$ and the default air temperature $u_{2,\text{default}}$ as proxies for energy consumption, following~\cite{balaji2013zonepac, chen2019gnu, bian2023bear}. 
\begin{equation}\label{eq:energy}
    Energy(t) = w_1 \|u_{1,t}\|_1 + w_2 \|u_{2,t} - u_{2, \text{default}}\|_1 
\end{equation}
$w_1, w_2$ are weight factors that vary with building types, and we set $u_{2, \text{default}} = 14.4^\circ$C according to~\cite{bian2023bear}. Other objective functions (e.g., time-of-use electricity price, peak demand charge) can also be flexibly included in the control problem based on the building system operation criteria.}

%% file: method.tex
\begin{figure*}[t]
    \centering
    \includegraphics[width=0.9\textwidth]{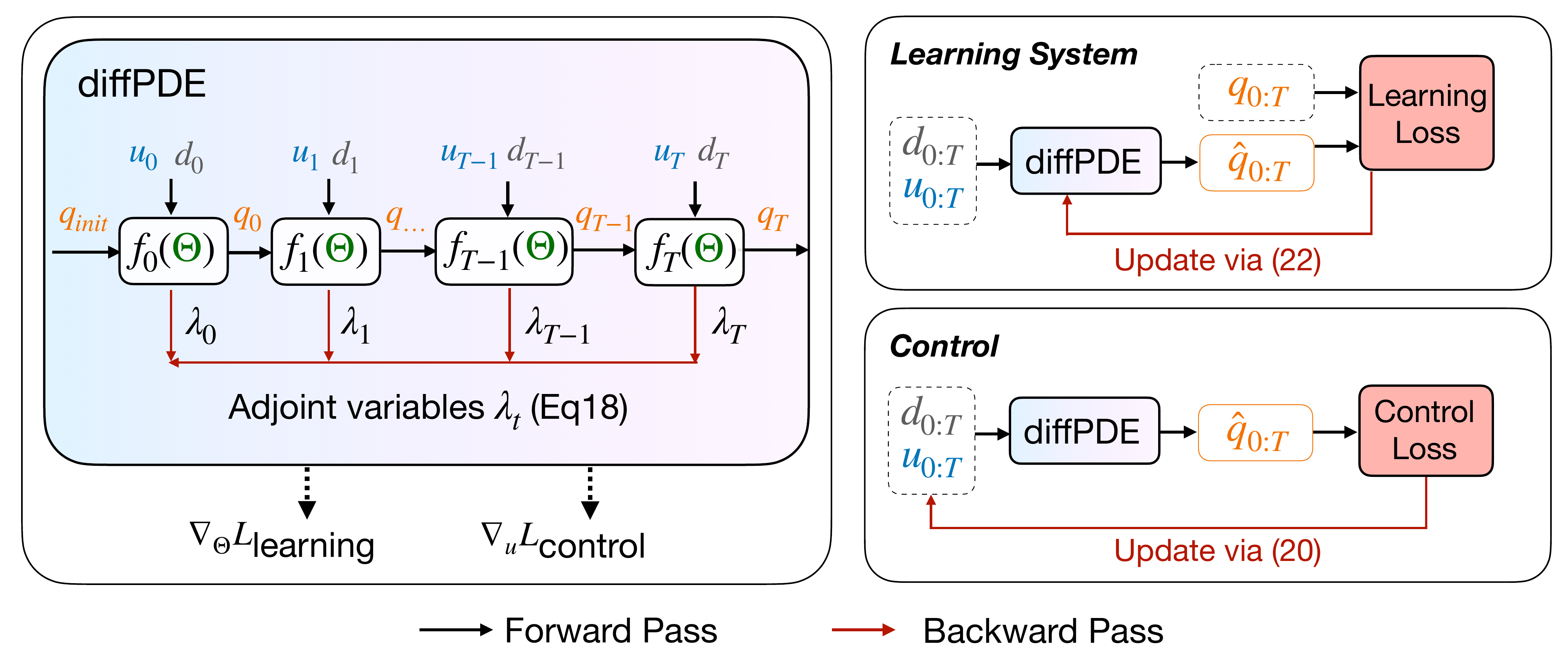}
    \caption{Algorithm for the PDE-based learning and control framework. For the ``diffPDE'' block, ``diff'' signifies ``differentiable'', denoting that it allows the computation of gradients. 
    Left figure shows how to derive the gradients: we first solve the PDEs in a forward pass, then we solve the adjoint variables $\lambda_t, t = 0, ..., T$ in a backward pass with~\eqref{eq:adjoint}. The gradient of model learning loss w.r.t. system parameters, and the gradient of control loss w.r.t. control actions are computed using~\eqref{eq:gradl} and~\eqref{eq:grad_control}, using the adjoint variables. Right figure illustrates updates of the model parameters and control via the obtained gradients. 
    }
    \label{fig:algorithm}
\end{figure*}
 We propose a gradient descent method for solving the system learning problem~\eqref{eq:p_sysid} and HVAC control problem~\eqref{eq:p_control}. We introduce the solution approach, gradient computation via a general PDE-constrained optimization formulation, and explain the essence of gradient computation via the adjoint method~\cite{mcnamara2004fluid}. An overview of the overall solution algorithm is illustrated in Figure~\ref{fig:algorithm} and will be described in detail at the end of section 4.1.

\subsection{Adjoint methods for PDE-constrained optimization} 
We consider the HVAC control problem in \eqref{eq:p_control} as an illustrative example for explaining the proposed solution method for solving PDE-constrained optimization.
Let us introduce a short-hand notation to represent the PDE dynamics in \eqref{eq:ns}-\eqref{eq:tc},
$$\frac{\partial q_t}{\partial t} = \mathcal{P}\left(q_t, \frac{\partial q_t}{\partial z}, \frac{\partial^2 q_t}{\partial z^2},  u_t, d_t; \Theta\right)$$
where $u_t$ specifies the control action at time $t$, and the term $q_t = [\vec{v}(z,t), C(z,t), T_e(z,t)]$ denotes the system state at time $t$. The term $d_t = [g(z,t), T_{\text{ambient}}(t)]$ represents the disturbance including occupancy and ambient temperature. $\mathcal{P}$ models the physical behavior of the PDE system defined in \eqref{eq:ns}-\eqref{eq:tc}, for a given set of system parameters $\Theta$. 

We consider a temporal discretization interval $\Delta t$, with the total number of discretization steps as $T = \frac{\overline{T}}{\Delta t}$. We also denote the initial system state as $q_{\text{init}}=q_{-1}$. 
The discrete-time version of optimal control problem~\eqref{eq:p_control} with the log barrier functions for the state constraints is written as, 
\begin{subequations} \label{eq:problem_d}
\begin{align}
    &\min_{u_0, u_1, \ldots, u_T} \quad  L_{\text{control}}:= \sum_{t=0}^T l(q_t, u_t) \label{eq:problem_d_obj}\\
    &\text{s.t.} \quad q_{t} = q_{t-1} + \Delta t \cdot \mathcal{P}\left(q_{t-1}, \frac{\partial q_{t-1}}{\partial z}, \frac{\partial^2 q_{t-1}}{\partial z^2}, u_t, d_t; \Theta \right)\,, \label{eq:discrete_dyn}\\
    & \quad \quad ~\eqref{eq:boundary_v}~\eqref{eq:boundary_t}~\eqref{eq:boundary_c} \tag{\text{Boundary conditions}} \\
    & \quad \quad \ \ \underline{u} \leq u_{t} \leq \overline{u}\,, \quad t = 0, ..., T.
    \label{eq:control_constr_discrete}
\end{align}
\end{subequations}
\revise{where $l(q_t, u_t):= \bar{l}(q_t, u_t) \Delta t$, with $\bar{l}(q_t, u_t)$ as the integral term in \eqref{eq:control_cost} (including the control cost and the barrier function terms)}.
Equation \eqref{eq:discrete_dyn} describes the discrete-time system evolution using the first-order Euler method. $\underline{u} = [\underline{u_1}, \underline{u_2}]^\tr, \overline{u} = [\overline{u_1}, \overline{u_2}]^\tr$. For further simplicity, we define the right hand side of \eqref{eq:discrete_dyn} as $f_t$, thus it can be re-written as,
\begin{equation}\label{eq:ft}
\begin{aligned}
        q_t & = f_t\left(q_{t-1}, u_t, d_t; \Theta\right).
\end{aligned}
\end{equation}
\revise{We plan to use a projected gradient descent method to solve the constrained optimization problem in~\eqref{eq:problem_d}. One critical challenge is the computation of gradients, namely $\nabla_{u_t} L_{\text{control}}\,, \forall t = 0, ..., T$ since all the state and control variables are restrained by the system dynamics equation \eqref{eq:ft}.
To overcome this challenge, we leverage the adjoint method, a well-established technique for PDE-constrained optimization~\cite{mcnamara2004fluid,lions1971optimal} that uses adjoint variables to enforce these dynamics constraints, then applying a projection operator to enforce the control constraints \eqref{eq:control_constr_discrete}.
}

Taking the total derivatives of both sides of~\eqref{eq:problem_d_obj}, 
\begin{equation}\label{eq:d_J}
    \delta \J = \sum_{t=0}^T \left(\frac{\partial l}{\partial q_t}  \delta q_t + \frac{\partial l}{\partial u_t} \delta u_t \right).
\end{equation}
Then, by taking the total derivative of equation~\eqref{eq:ft} and re-arrange the equation, we get
\begin{equation}\label{eq:d_f}
    \delta q_{t} - \frac{\partial f_t}{\partial q_{t-1}} \delta q_{t-1} - \frac{\partial f_t}{\partial u_t}  \delta u_t - \frac{\partial f_t}{\partial d_t}  \delta d_t= 0.
\end{equation}

\revise{Note that the left side of equation~\eqref{eq:d_f} is always zero. The adjoint method multiplies the adjoint variables $\lambda_t$ by the left side of equation~\eqref{eq:d_f} and adds to equation~\eqref{eq:d_J}. As a result, we obtain}
\begin{equation*}\label{eq:d_all}
    \begin{aligned}
        \delta \J & = \sum_{t=0}^{T} \left(\frac{\partial l}{\partial q_t}\delta q_t + \frac{\partial l}{\partial u_t} \delta u_t \right)+ \\
        & \quad \sum_{t=0}^{T} \lambda_{t}^\tr \left(\delta q_{t} - \frac{\partial f_t}{\partial q_{t-1}} \delta q_{t-1} - \frac{\partial f_t}{\partial u_t}  \delta u_t - \frac{\partial f_t}{\partial d_t}  \delta d_t \right) \\
        & = \left(\frac{\partial l}{\partial q_{T}} + \lambda_{T}^\tr\right) \delta q_{T} +  \sum_{t=0}^T \left(\frac{\partial l}{\partial u_t}-\lambda_t^\tr \frac{\partial f_t}{\partial u_t} \right)\delta u_t   \\
        & \quad \,
        +\sum_{t=0}^{T-1} \left(\frac{\partial l}{\partial q_t}+\lambda_t^\tr - \lambda_{t+1}^\tr \frac{\partial f_{t+1}}{\partial q_t}\right)\delta q_t
         \\
        & \quad \, - \lambda_0^\tr \frac{\partial f_0}{\partial q_{-1}}\delta q_{-1} - \sum_{t=0}^T \lambda_t^\tr \frac{\partial f_t}{\partial d_t} \delta d_t.
    \end{aligned}
\end{equation*}
By dividing both side with $\delta u_t$, we get the gradient of control loss $\J$ with respective to (w.r.t.) the control action $u_t$, 
\begin{equation}\label{eq:d_all2}
    \begin{aligned}
    \frac{\delta \J}{\delta u_t} & = \left(\frac{\partial l}{\partial q_{T}} + \lambda_{T}^\tr\right) \underbrace{\frac{\delta q_{T}}{\delta u_t}} +  \sum_{t=0}^T \left(\frac{\partial l}{\partial u_t}-\lambda_t^\tr \frac{\partial f_t}{\partial u_t}\right) \underbrace{\frac{\delta u_t}{\delta u_t}} \\
    & \quad \,
 +    \sum_{t=0}^{T-1} \left(\frac{\partial l}{\partial q_t}+\lambda_t^\tr - \lambda_{t+1}^\tr \frac{\partial f_{t+1}}{\partial q_t}\right) \underbrace{\frac{\delta q_t}{\delta u_t}}\\
    & \quad \, - \lambda_0^\tr \frac{\partial f_0}{\partial q_{-1}} \underbrace{\frac{\delta q_{-1}}{\delta u_t}} - \sum_{t=0}^T \lambda_t^\tr \frac{\partial f_t}{\partial d_t} \underbrace{\frac{\delta d_t}{\delta u_t}}.
    \end{aligned}
\end{equation}
\revise{The five terms, each highlighted with an underbrace, correspond to gradients relevant to our system analysis: the gradient of final state $q_T$ w.r.t. $u_t$, the gradient of control action $u_t$ w.r.t. itself, the system state at each time step $q_t$ w.r.t. $u_t$, the gradient of initial system state $q_{-1}$ w.r.t $u_t$, and the gradient of disturbance $d_t$ w.r.t $u_t$. We have $\delta u_t / \delta u_t$ = 1, $\frac{\delta d_t}{\delta u_t} = 0$ and $\frac{\delta q_{-1}}{\delta u_t} = 0$, as the gradient of a variable w.r.t. itself is unity, and the initial system state and disturbance are not affected by control actions. }

To eliminate the terms $\frac{\delta q_t}{\delta u_t}$ for $t=0, \ldots, T$ for the ease of computaion, the adjoint method sets the adjoint variables $\lambda_t$ backwards as follows,
\begin{subequations}
\label{eq:adjoint}
\begin{align}
       \lambda_T &= -\frac{\partial l}{\partial q_T}^\tr, \\
       \lambda_t &= \left(\frac{\partial f_{t+1}}{\partial q_t}\right)^\tr \lambda_{t+1} - \left(\frac{\partial l}{\partial q_t}\right)^\tr, t = 0, \ldots, T-1,
\end{align}
\end{subequations}

\revise{Thus, by plugging in the adjoint variable values satisfying \eqref{eq:adjoint}, and setting $\frac{\delta d_t}{\delta u_t} = \frac{\delta q_{-1}}{\delta u_t} = 0$ and $\frac{\delta u_t}{\delta u_t} = 1$, the desired gradient of the control loss $L_{\text{control}}$ w.r.t. the control variables $u_t$ can be computed as follows,}
\begin{equation}\label{eq:grad_control}
   \nabla_{u_t} L_{\text{control}} = \frac{\delta L_{\text{control}}}{\delta u_t} = \sum_{t=0}^{T} (\frac{\partial l}{\partial u_t} -\lambda_{t}^\tr \frac{\partial f_t}{\partial u_t}).
\end{equation}
\revise{We are now able to optimize the control variables with the obtained the gradient:}
\begin{subequations}\label{eq:control_update}
\begin{align}
    & u_t \leftarrow u_t - \eta \cdot  \nabla_{u_t} L_{\text{control}}, \\
    & u_t \leftarrow {\rm Proj}(u_t, \underline{u}, \overline{u}). \label{eq:project}
\end{align}
\end{subequations}
with the projection operator ${\rm Proj}(x, a, b)$ defined as,
\begin{eqnarray}
            {\rm Proj}(x, a, b) := 
            \begin{cases}
              a, &\mbox{ if } x < a \\
              b, &\mbox{ if } x > b \\
              x, &\mbox{ else }
            \end{cases}
  \label{eq:proj_def}
\end{eqnarray}
\revise{Here, $\eta$ represents the learning rate, and equation~\eqref{eq:project} describes the projected gradient descent.
}

For the learning task, the decision variable to the optimization problem changes to the system parameter $\Theta = \{\nu, k_{T_e}, k_C, \alpha, \alpha_{T_e}, \alpha_C\}$. We update the system parameter estimates as follows, 
\begin{subequations}
\label{eq:learning_update}
\begin{align}
  &  \Theta \leftarrow \Theta - \eta \cdot \nabla_{\Theta} L_{\text{learning}}, \label{eq:gradl}\\
 &   \nabla_{\Theta} L_{\text{learning}} = \sum_{t=0}^{T} -\lambda_{t}^\tr \frac{\partial f_t}{\partial \Theta}, \label{eq:grad_learning}
\end{align}
\end{subequations}
where the adjoint variable is computed using~\eqref{eq:adjoint}. 

\revise{We illustrate the solution approach in Figure~\ref{fig:algorithm}.  The left figure illustrates the derivation of gradients based on the adjoint method. We first solve the PDEs in the forward pass. At each time step $t = 0, 1, \ldots, T$, we solve the PDEs symbolized by $f_t$ to obtain the state for the next time step $q_{t} = \{\vec{v}(z,t), T_e(z,t), C(z,t)\}$. When the PDE operations are implemented in a differentiable manner, the automatic differentiation tools~\cite{holl2020phiflow,paszke2017automatic} can chain the derivatives of these operations with built-in machine learning operations to compute the analytic derivatives $\frac{\partial f_t}{\partial q_t}$, $\frac{\partial f_t}{\partial  u_t}$, and $\frac{\partial f_t}{\partial \Theta}$ accordingly. The adjoint variables $\lambda_T, \lambda_{T-1}, \ldots, \lambda_0$ in~\eqref{eq:adjoint} can be computed in a backward pass. Consequently, we can compute the learning loss, control loss, the gradients of the learning loss w.r.t. system parameters $\nabla_\Theta L_{\text{learning}}$, and the gradients of the control loss w.r.t. control actions $\nabla_{u} \J$. 
The right figure illustrates the updates to system parameters and control variables using gradients for the learning and control problems.
In our implementation, we built our differentiable solver based on the PhiFlow framework~\cite{holl2020phiflow}.}



\subsection{Learning and control algorithms}
The system learning procedure and control procedure are summarized in Algorithm~\ref{alg:learning} and Algorithm~\ref{alg:control}, respectively. 

For learning system phase (Algorithm~\ref{alg:learning}), we require the offline dataset with recordings $g(z,t), T_{\text{ambient}}(t)$, 
the history control sequences $u_t$, and measured CO$_2$ concentrations and temperatures $C(z_{sensor},t), T_e(z_{sensor},t)$. At each iteration $k$, we solve the PDEs based on estimated building parameter $\Theta^{(k)}$, and update it following the update law in \eqref{eq:learning_update} to minimize the learning loss.

\revise{Once Algorithm~\ref{alg:learning} has estimated the system parameters $\Theta$, we are then able to optimize the control variables using these estimated parameters. For control phase (Algorithm~\ref{alg:control}), we require the initial system state $q_{init}=q_{-1}$, predicted disturbance $g(z,t), T_{\text{ambient}}(t)$ and estimated system parameters $\Theta$. We first initialize the control action sequence $u^{(0)}$. At each iteration $k$, we solve the PDEs based on control actions $[u_0^{(k)},...,u_T^{(k)}]$ and update the actions following the update law in \eqref{eq:control_update} to minimize the control loss.}

\begin{algorithm}[ht]
\caption{Algorithm for \textcolor{teal}{Learning System} Phase}\label{alg:learning}
\begin{algorithmic}[1]
\Require Dataset $D = \{g(z,t), T_{\text{ambient}}(t), u_t$, $C(z_{sensor},t),T_e(z_{sensor},t)$, $t=0, \ldots, T\}$.
\Ensure $\Theta^{(0)} = \{\nu^{(0)}, k_{T_e}^{(0)}, k_C^{(0)}, \alpha^{(0)}, \alpha_{T_e}^{(0)}, \alpha_C^{(0)}\}$ \Comment{initial parameters}
\For{$k = 0, 1, \ldots, K$}
\State \textbf{sample} data $B = \{(C(z_{sensor},t),T_e(z_{sensor},t)), t = 0, \ldots, T\}$ from dataset $D$
\State \textbf{compute} the airflow velocity $\widehat{\vec{v}}(z,t)$ for $t = 0, \ldots, T$  using~\eqref{eq:ns} with $\nu^{(k)}$
\State \textbf{compute} the predicted temperature $\widehat{T}_e(z,t)$, and predicted CO$_2$ $\widehat{C}(z,t)$ for $t=0, \ldots, T$ using~\eqref{eq:tc} with  $k_{T_e}^{(k)}, k_C^{(k)}, \alpha^{(k)}, \alpha_{T_e}^{(k)}, \alpha_C^{(k)}$ 
\State \textbf{evaluate} the learning loss function via~\eqref{eq:l_cost1}\eqref{eq:l_cost2}
\State \textbf{update} system parameters $\Theta^{(k)}$ with $\nabla_{\Theta^{(k)}} L_{\text{learning}}$:
$$\Theta^{(k+1)} \leftarrow \Theta^{(k)} - \eta \cdot \nabla_{\Theta^{(k)}}L_{\text{learning}}$$
\EndFor
\State \textbf{Return $\Theta$}
\end{algorithmic}
\end{algorithm}
\begin{algorithm}[ht]
\caption{Algorithm for \textcolor{teal}{Control} Phase}\label{alg:control}
\begin{algorithmic}
\Require Initial state variable $q_{init}=q_{-1}$, predicted disturbance $g(z,t), T_{\text{ambient}}(t), t = 0, \ldots, T$ and learned parameters $\Theta$
\Ensure $u^{(0)}=[u^{(0)}_0, \ldots, u^{(0)}_T]$ \Comment{initial control variables}
\For{$k = 0, 1, \ldots, K$}
\State \textbf{compute} the airflow velocity $\widehat{\vec{v}}(z, t), t = 0, \ldots, T$ using~\eqref{eq:ns} with $u^{(k)}$
\State \textbf{compute} the predicted temperature $\widehat{T}_e(z,t)$, and predicted CO$_2$ $\widehat{C}(z,t)$ for $t = 0, \ldots, T$ using~\eqref{eq:tc} with $u^{(k)}$
\State \textbf{evaluate} the control loss function via~\eqref{eq:control_cost}
\State \textbf{update} control variables $u^{(k)}$ following the update law in~\eqref{eq:control_update}:
$$u^{(k+1)} \leftarrow  u^{(k)} - \eta \cdot \nabla_{u^{(k)}} \J$$
$$u^{(k+1)} \leftarrow {\rm Proj}(u^{(k+1)}, \underline{u}, \overline{u})$$
\EndFor
\State \textbf{Return $u$}
\end{algorithmic}
\end{algorithm}

%% file: experiment.tex
In this section, we present the capabilities of the proposed framework in two key tasks: system learning and optimal HVAC control. We illustrate how our framework can accurately learn the unknown system parameters using historical data.  Additionally, we demonstrate the performance of our method in the HVAC control, where it significantly outperforms existing control methods including maximum airflow policy, and learning-based control with reinforcement learning~\cite{schulman2017proximal} and optimization-based control with ODE models~\cite{li2022tube, zhang2021novel, bian2023bear, xiao2023building}.
The source code, input data, and trained models from all experiments will be available on GitHub\footnote{\url{https://github.com/alwaysbyx/PDE-HVAC-control}}.


We build the testbed shown in Figure~\ref{fig:boundary}. For all the simulation unless otherwise specified, we consider a room with the length $z_x = 4.2$m, and the height $z_y = 2.7$m, which has dimensions identical to the conference room discussed in~\cite{jin2015sensing}. This conference room bears close resemblance to other typical indoor spaces, with a ventilation system including air supply and air return vents on the ceiling.
The computational mesh for this space is defined by the size tuple $(n_x,n_y)=(42,27)$ where $n_x$ and $n_y$ represent the number of discrete divisions along the room's length and height, respectively. The PDE discretization time step $\Delta t$ is chosen as 1 minute. 
The occupancy position are $\{z=(x,y)|2.5\leq x \leq  3.1, 0.7\leq y\leq 0.9\}$.
Overall, the parameters for all the experiments are listed in Table~\ref{tab:exp}. \revise{We follow works~\cite{bian2023bear, he2016zoned} to select the appropriate maximum and minimum airflow rates for the considered room.}
\begin{table}[h]
    \centering
    \caption{Parameters for the experiments. }
    \begin{tabular}{ll}\hline
   Notation     &  Value \\\hline
    $\Z_{\text{supply}}$ & $\{z=(x,y)|2.7\leq x \leq 3.3, y=2.7\}$\\
    $\Z_{\text{return}}$ & $\{z=(x,y)|0.9\leq x \leq 1.5, y=2.7\}$\\
    $\Z_{\text{outside}}$ & $\{z=(x,y)|x=4.2, 0\leq y\leq 2.7\}$\\ 
    $\overline{u_1}$  & maximum airflow rate: 1.2\,m/s \\
    $\underline{u_1}$  & minimum airflow rate: 0.12\,m/s\\
    $\Delta t$ & 1 minute \\
    \hline
    \end{tabular}
    \label{tab:exp}
\end{table}


\begin{table}[t]
    \footnotesize
    \centering
    \caption{Model parameter learning for the joint temperature and CO$_2$ field learning. 
    }
    \begin{tabular}{lll}
    \hline 
      Parameter   & True value & Estimated value \\ \hline
      $\nu$ & 0.00108 & 0.00106  \\
        $k_C$ & 0.00108 & 0.00106  \\
        $k_{T_e}$ & 0.002 & 0.002  \\
         $\alpha$ & 0.65 & 0.048  \\
          $\alpha_C$ & 0.83 & 0.83  \\
           $\alpha_{T_e} $ & 0.00500 & 0.00496  \\
      \hline
    \end{tabular}
    \label{tab:learn_sythetic}
\end{table}

\begin{table}[h]
    \footnotesize
    \centering
    \caption{Sensor Placement $z_{sensor}$. 
    }
    \begin{tabular}{ll}
    \hline 
      Position Description   & Position values \\\hline
      Ground &  (0.1m, 0m), (3m, 0m) \\
      Supply vent & (3m, 2.7m) \\
      Return vent & (0.9m, 2.7m) \\
      Wall & (0.1m, 0.7m) \\
      Table & (3m, 0.8m) \\
      \hline
    \end{tabular}
    \label{tab:sensor_sythetic}
\end{table}


\begin{figure*}[t]
    \centering
    \includegraphics[width=1.7\columnwidth]{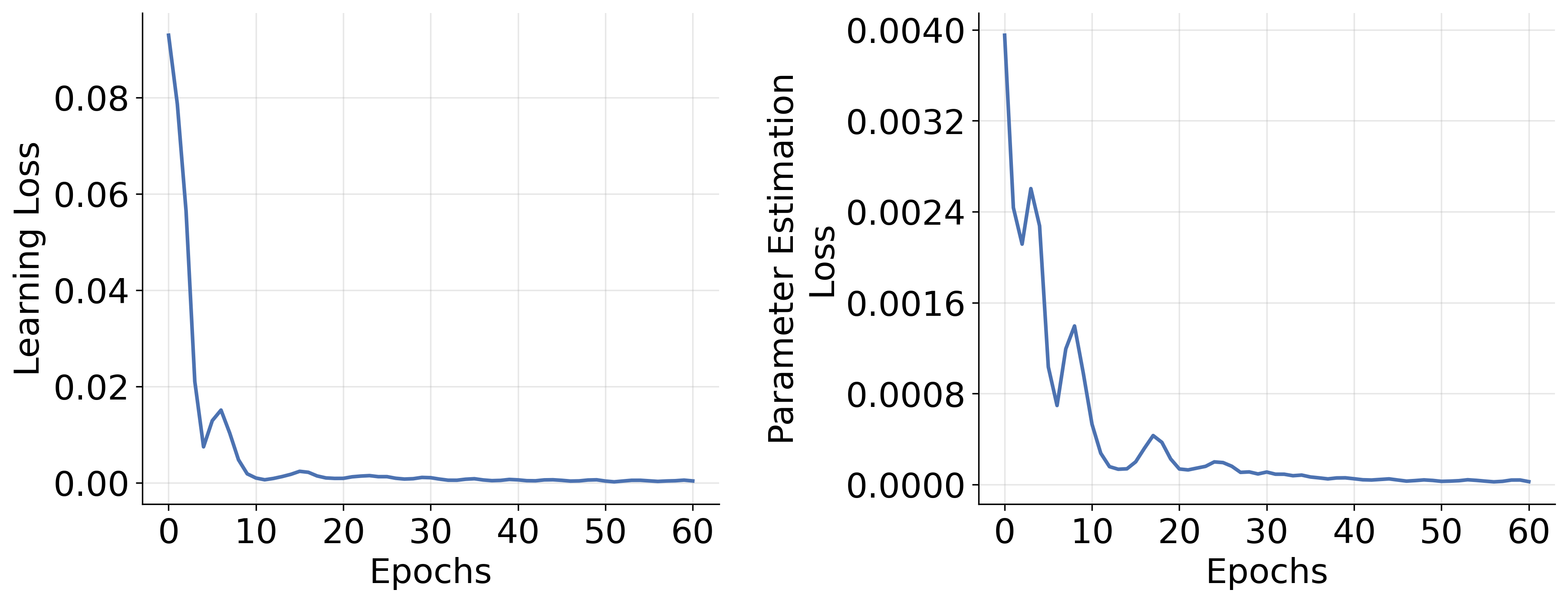}
    \caption{Convergence results for joint temperature and CO$_2$ field experiment: the learning loss and the parameter estimation loss curves.}
    \label{fig:learn_synthetic}
\end{figure*}
\begin{figure*}[t]
    \centering
    \includegraphics[width=1.85\columnwidth]{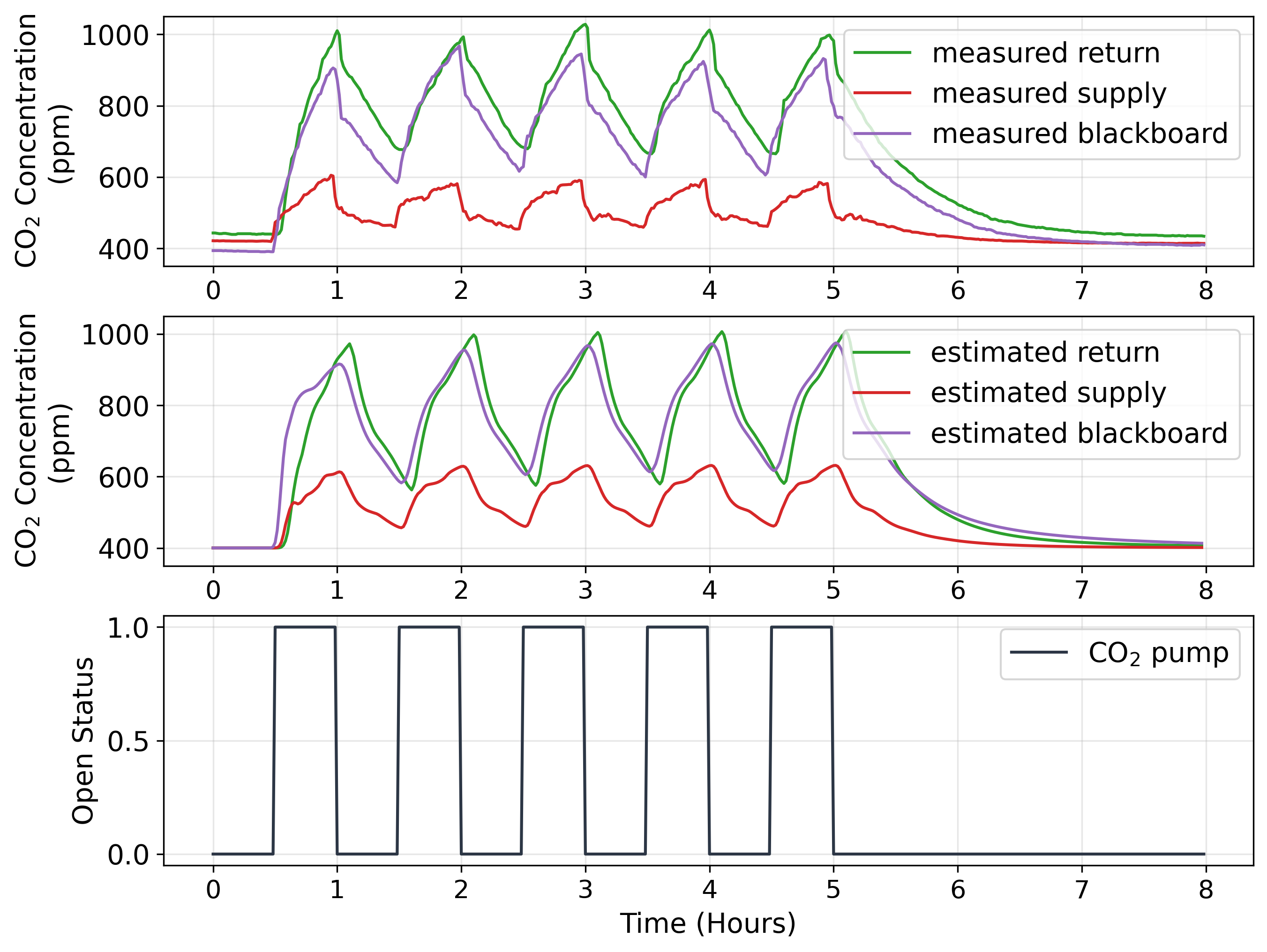}
    \caption{Comparison of measured CO$_2$ concentrations from a real world dataset~\cite{jin2015sensing} (top) and estimated CO$_2$ concentrations (middle) based on the learned PDE model. The bottom figure represents the open status of the CO$_2$ pump in the real world experiment. An open CO$_2$ pump (indicated by a value of 1) corresponds to $g(z,t)=1$, while a closed CO$_2$ pump (indicated by a value of 0) corresponds to $g(z,t)=0$.
    }
    \label{fig:learn_CO$_2$}
\end{figure*}
\subsection{System model learning}
\subsubsection{Case 1. joint temperature and CO$_2$ field learning}\label{sec:synthetic}
Here, the model parameters are $\Theta = \{\nu, k_C, k_{T_e}, \alpha, \alpha_C, \alpha_{T_e}\}$. For the model parameters, we have chosen specific values as presented in Table~\ref{tab:learn_sythetic}.
The selection of these values follows works~\cite{farahmand2017deep,kreith1999fluid,linden1999fluid}.
The fan speed is set to be consistent during the simulation, $\vec{v}(z,t) = - \overline{u}_1 \cdot e_y, \forall z \in \Z_{\text{supply}}$. 
The simulation runs for a duration of $\overline{T}=60$ minutes. 
The sensor placement is listed in Table~\ref{tab:sensor_sythetic}.
The training process is conducted over 60 epochs. The learning loss of the joint temperature and CO$_2$ field learning is shown in Fig~\ref{fig:learn_synthetic} (left). We observe that the learning loss is decreasing as the epoch increases. \revise{Additionally, we visualize the curve of parameter estimation loss, which is defined as the mean squared error between the predicted parameters $\hat{\Theta}$ and the actual parameters $\Theta$:
\begin{equation}
    \text{Parameter Estimation Loss} = \|\Theta - \hat{\Theta}\|^2.
\end{equation}}
Consistent with the learning loss, the parameter estimation loss also shows a reduction as the number of epochs increases. True model parameters and estimated parameters are listed in Table~\ref{tab:learn_sythetic}, which indicates that the system model can be learned precisely with low error.

\subsubsection{Case 2. a real world dataset for CO$_2$ field learning}
We now test the performance of the proposed approach using a real-world dataset in~\cite{jin2015sensing}. The dataset is collected in a conference room on the UC berkeley campus, which shares the same dimensions and ventilation system as depicted in Figure~\ref{fig:boundary}.
The sensors are placed on both vents, in addition to the blackboard on the sidewall to sense CO$_2$ concentrations. \revise{The sensor locations for the return vent are set as $\{(x=1m,y=2.7m),(x=0.9m,y=2.7m)\}$, while the position for the supply vent sensor is set at $\{x=3.0m, y=2.7m)\}$. The area near the blackboard is defined by points $\{(x=0.1m,y=0.7m),(x=0.2m,y=0.8m)\}$. In instances where the measurement points exceed one, we calculate the average of these values to represent the CO$_2$ concentration at specific positions, which serves to provide a more accurate and representative measurement of the CO$_2$ levels. 
}

\revise{Occupancy was simulated via a CO$_2$ pump, and the CO$_2$ pump was operated periodically, being switched on for 30 minutes and then turned off for 30 minutes. The CO$_2$ pump position is set as the same as the occupancy position. The simulation runs for a duration of $\overline{T}$ = 480 minutes. Figure~\ref{fig:learn_CO$_2$}  (bottom) shows the operation status during the experiment, when the CO$_2$ pump is open (indicated by a value of 1), $g(z,t)=1$, and when the CO$_2$ is off (indicated by a value of 0), $g(z,t)=0$. }
Figure~\ref{fig:learn_CO$_2$}(top) illustrates the CO$_2$ measurements at the supply vent, return vent, and blackboard. We observe that the measurements exhibit periodic patterns driven by the on/off status of the CO$_2$ pump, and the concentration varies both spatially and temporally. For instance, the CO$_2$ concentration at blackboard will increase before CO$_2$ concentration at return vent increases. In addition, 
the air supply vent has a smaller magnitude of CO$_2$ concentration compared to the blackboard and the air return vent. When there is no CO$_2$ releasing, the CO$_2$ concentrations stabilize and reach a plateau.

For this real-world dataset, the learning parameter is a subset of $\Theta$ as $\{\nu, k_C, \alpha, \alpha_C\}$ since there is no temperature measurement available for estimating $\{k_{T_e}, \alpha_{T_e}\}$. For the loss function, we employ the mean-squared error between the actual CO$_2$ measurements and predicted values of CO$_2$ at the sensor locations (supply vent, return vent, and blackboard), $L(\Theta) =  \sum_{t=0}^T  \|C(z_{sensor,t}) - \widehat{C}(z_{sensor},t)\|^2$.
    

\revise{The estimated CO$_2$ concentrations based on the learned model parameters are shown in Figure~\ref{fig:learn_CO$_2$} (middle). }
The experiment result demonstrates that the proposed method can capture CO$_2$ concentration patterns. Specifically, we observe that the CO$_2$ concentration at the return vent consistently showed the highest magnitude, while the concentration at the supply vent was the lowest. Additionally, CO$_2$ levels near the blackboard demonstrated a tendency to rise and fall sooner than those at the return vent. The mean absolute percentage error (MAPE) between the measured CO$_2$ concentrations and the estimated CO$_2$ concentrations is 6.89\%. This demonstrates the applicability and robustness of our method in dealing with real-world data with potential measurement noise and incomplete information.

\begin{figure*}[ht]
    \centering
    \includegraphics[width=1.85\columnwidth]{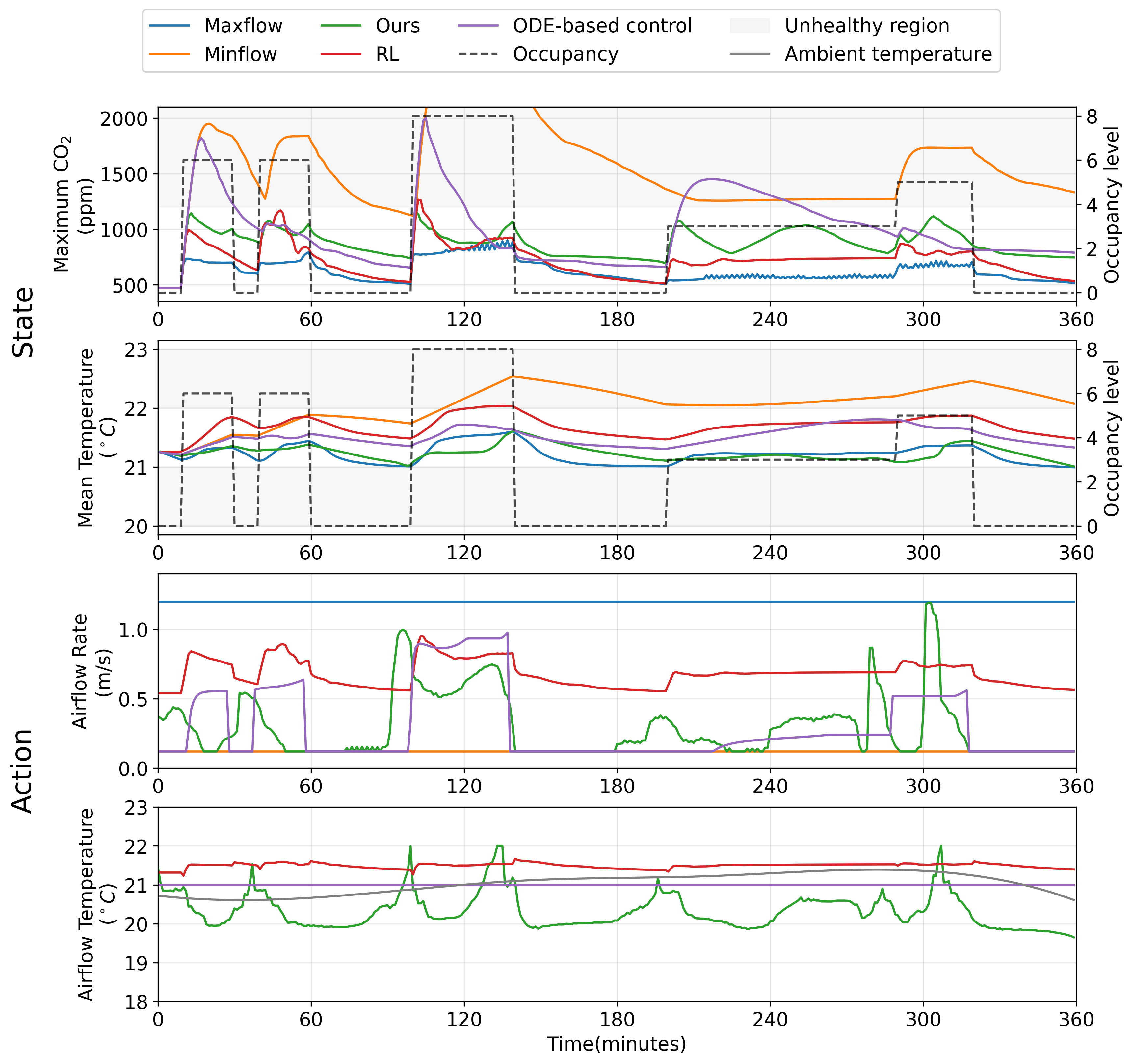}
    \caption{Comparison of building control performance using our methods and baselines, including Maxflow control, \revise{Minflow control, optimization-based control with ODE models (ODE-based control)}, and reinforcement learning (RL). The figure shows one example of control actions and indoor temperature and CO$_2$ dynamics. \revise{In the top two figures, black dashed line represents the occupancy and the grey area denotes the unhealthy region. In the bottom figure, grey line represents the ambient temperature.} }
    \label{fig:control}
\end{figure*}

\subsection{Building control}
We have demonstrated promising results in the task of learning unknown building parameters in the PDE models. Subsequently, we show how our framework can optimize 
HVAC control to reduce the energy consumption while ensuring health/comfort constraints. 

\subsubsection{Experiment setting} 
We assume that the building management system has learned the building model from historical data and is now focused on solving the control problem~\eqref{eq:p_control}. The model parameters are established as described in Section~\ref{sec:synthetic}. 
In the control task, the management system determines the control action $\{(u_{1,t}, u_{2,t}), t\in [0, \ldots, T]\}$, where $u_{1,t}$ is the supply airflow rate (m/s), and $u_{2,t}$ indicates the temperature ($^\circ$C) of the air supplied to the room. 
The CO$_2$ limit is set to be $C_{\text{max}} = 1200\text{ppm}, \forall z, t$, \revise{recommended by~\cite{zhang2023model} as the maximum CO$_2$ level for human health.} 
\revise{The temperature limits are set to a rather tight range as $T_{\text{min}}=21.0^\circ C$ and $T_{\text{max}} = 22.0^\circ C$, for testing the controller ability in maintaining thermal comforts. The values are chosen based on~\cite{chen2019gnu}, where the zone temperature generally remains between 21$^\circ C$ to 22$^\circ C$ under the existing building controller.} 
\revise{The simulation runs for a duration of $\overline{T}=360$ minutes.} Specifically, we utilize data corresponding to San Francisco's temperature in \revise{12pm to 6pm on July 1, 2023}. 
We assume the building occupancy schedule and the ambient temperature profiles are known. 
\revise{Without generalization, we set the following parameters in the control cost function as $\alpha_1 = 0.1, \alpha_2 = 0.2, \alpha_3 = 0.5$, $w_1 = 30, w_2=3$.}
\begin{table*}[b]
\centering
\caption{HVAC Control: Performance of the proposed approach in comparison with baselines, including Maxflow control, \revise{Minflow control, optimization-based control with ODE models (ODE-based control)}, and reinforcement learning (RL). 
}
\begin{tabular}{llllll}
\hline
Method                 & Energy Consumption (kWh) & \multicolumn{2}{l}{Temperature violation ($^\circ$C)} & \multicolumn{2}{l}{CO$_2$ violation (ppm)} \\
                       &                    & average             & maximum             & average         & maximum         \\ \hline
Maxflow control        & 334.1        & 0                   & 0                   & 0               & 0               \\
Minflow control            & 139.7   & 0.16               & 0.54              & 377.1          & 1164.2          \\
ODE-based control  & 172.0 & 0.0 & 0.0 & 57.0   & 800.0   \\
RL & 249.1         & 0.001                   & 0.04                  & 0.35           & 65.6           \\
Ours                   & \textbf{158.4}   &  \textbf{0}                   &  \textbf{0}                   &  \textbf{0}               &  \textbf{0}               \\ \hline
\end{tabular}\label{tab:cost}
\end{table*}
\begin{figure}[t]
    \centering
    \includegraphics[width=0.95\columnwidth]{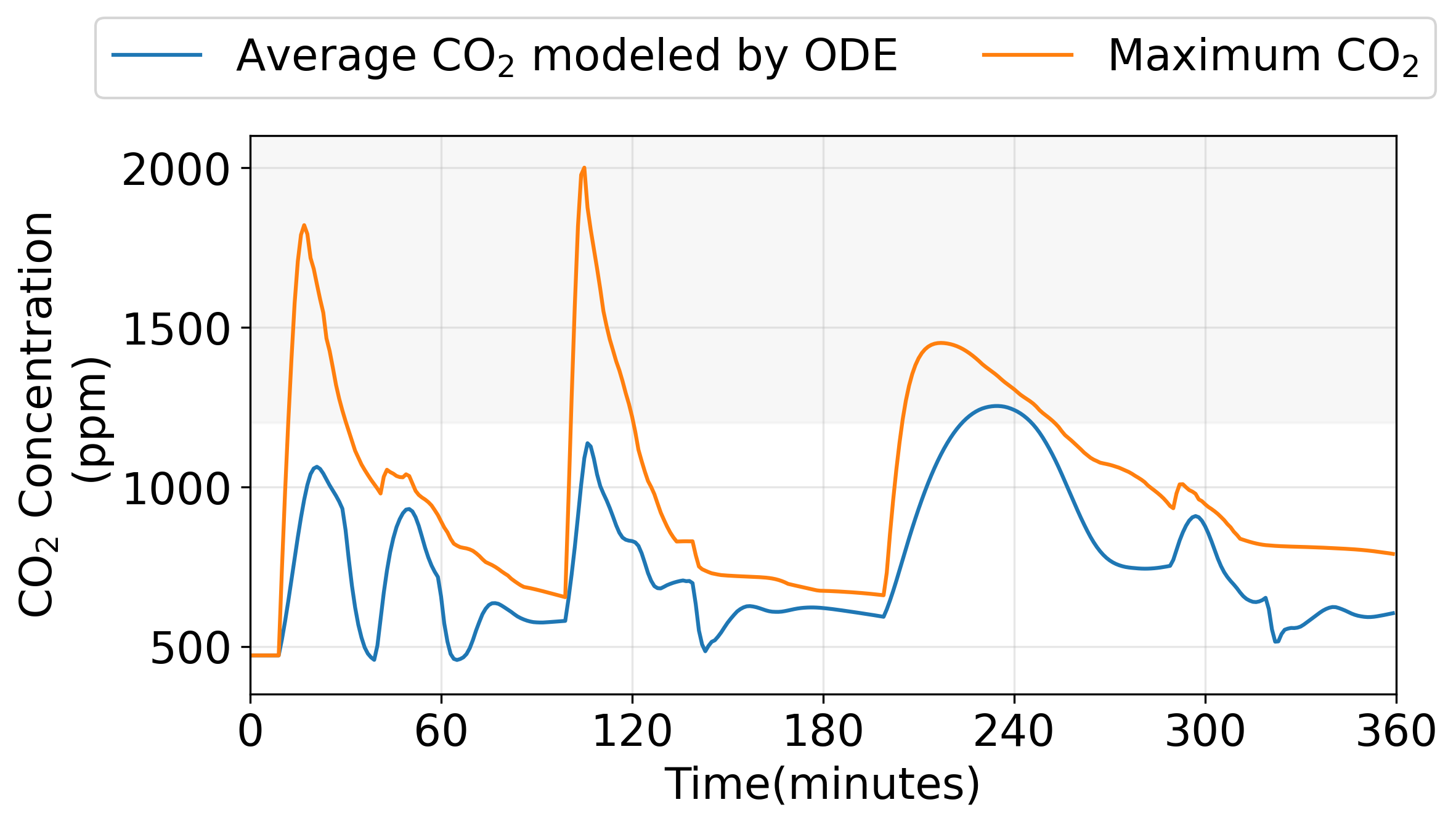}
    \caption{Average CO$_2$ concentration modeled by ODE (blue) and maximum CO$_2$ concentration (orange). Although CO$_2$ modeled by ODE generally stays within healthy limits, there are notable violations for the maximum CO$_2$ levels. }
    \label{fig:compare-ode}
\end{figure}

\subsubsection{Baselines}
We compare the proposed algorithm against a set of baselines that include both traditional control methods and cutting-edge, learning-based techniques. 
\revise{The traditional methods consist of a maximum airflow policy (Maxflow control)~\cite{bian2023bear} and a minimum airflow policy (Minflow control).}
Additionally, we explore optimization-based control with ODE models \revise{(ODE-based control)} and learning-based control with reinforcement learning (RL). The comparison of our approach with ODE-based control is motivated by the widespread use in current literature~\cite{li2022tube, zhang2021novel, bian2023bear, xiao2023building} to model thermal and CO$_2$ dynamics using ODEs, such as reduced linear Resistance-Capacitance (RC) models. However, these models often overlook spatial interrelationships, a gap our study aims to address. RL, recognized as a state-of-the-art technique, has been increasingly adopted in recent building HVAC control research~\cite{hosseinloo2023data,zhang2023bear,yu2020multi,wang2024energy, du2021intelligent}. Our comparison here aims to illustrate how our approach compares against different existing building control methods. 

We include a brief introduction of the bench-marking methods. Details of implementation for \revise{ODE-based control and RL-based control} are described in Appendix A and Appendix B, respectively.
\begin{itemize}
    \item \emph{Maxflow control:} operates at the \emph{maximum} supply airflow rate. To ensure that the temperature does not fall below the lower limit and to reduce the energy cost, this policy sets the supply air temperature to the minimum threshold, denoted as $u_{2,t} = T_{\text{min}} = 21^\circ C$.
    \item \emph{Minflow control:} operates at the \emph{minimum} supply airflow rate to minimize the energy cost associated with airflow rates. The policy also sets the supply air temperature to the minimum, i.e., $u_{2,t} = T_{\text{min}} = 21^\circ C$.
    \item \emph{Optimization-based control with ODE models (ODE-based control)}: 
    \revise{employs ODE models to model the average CO$_2$ in a focus area and the average temperature in a room. Following this,  After solving the ODE-constrained optimization problem, the control actions are implemented on the building PDE environment. }
    \item \emph{Reinforcement learning (RL)}: RL learns a control policy through direct interaction with the introduced building PDE environment. At each time step $t$, the RL agent selects an action $u_t$ given the current temperature $T_e(z,t), \forall z$ and CO$_2$ concentrations $C(z,t), \forall z$. 
\end{itemize}

\subsubsection{Control results} 
\begin{figure*}[t]
    \centering
    \includegraphics[width=2.0\columnwidth]{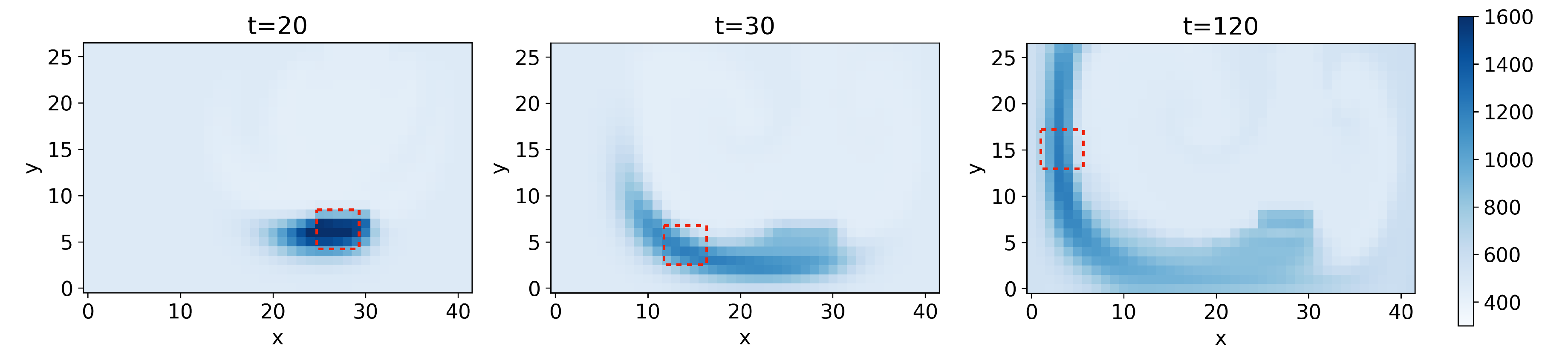}
    \caption{The distribution of CO$_2$ levels for different $t$ (minute) with the maximum CO$_2$ concentration marked by red dashed boxes. The locations of  the maximum CO$_2$ concentrations vary over time. }
    \label{fig:CO2}
\end{figure*}

Figure~\ref{fig:control} illustrates an example of control actions along with the dynamics of indoor temperature and CO$_2$. Table~\ref{tab:cost} presents performance associated with our proposed approach and baseline methods. Notably, the Minflow control policy results in significant violations of CO$_2$ and indoor temperature constraints. The Maxflow control, while ensuring a healthy environment by maximizing the airflow rate, leads to high energy consumption. Our proposed approach demonstrates a 52.6\% reduction in energy costs compared to the Maxflow policy. The RL approach manages to save energy but fails to meet temperature and CO$_2$ constraints.

Optimization-based control with ODE dynamic models performs better than RL, leading to lower energy usage. However, it neglects spatial variations of CO$_2$ distribution, and leads to violations of CO$_2$ constraint. Figure~\ref{fig:compare-ode} displays the curves for maximum CO$_2$ in a room and the CO$_2$ concentrations modeled with the ODE approach. While the modeled CO$_2$ levels generally remain within the healthy limits, the maximum CO$_2$ concentrations can occasionally exceed these limits. Additionally, Figure~\ref{fig:CO2} illustrates how CO$_2$ distributions change over time under ODE-based control. The maximum CO$_2$ levels are achieved in red dashed boxes and we observe that the locations of the maximum CO$_2$ can be different. 
The results demonstrate that the ODE approach, fails to capture the rich spatial-temporal dynamics of airflow velocity fields and ``dead zones'' of high CO$_2$/aerosol concentrations, thus can lead to violation of health constraints. In this way, ODE modeling is insufficient for effective building ventilation design.
In contrast, our approach not only adheres to all constraints but also shows a 7.9\% and 36.4\% energy consumption reduction compared to ODE-based control and RL, respectively.



%% file: appendix.tex
\subsection*{Appendix A. Implementation details of the optimization-based control with ODE models}\label{sec:app_a}
We model the thermal dynamics using a RC model~\cite{bian2023bear, xiao2023building}:
\begin{equation}\label{eq:ODE_t}
\begin{aligned}
       c \frac{d T_e(t)}{d t} & = \frac{T_{\text{ambient}}(t) - T_e(t)}{R} \\
       & + c_p u_{1,t} (u_{2,t} - T_e(t)) + p_g g(t)
\end{aligned}
\end{equation}
where $c$ is heat capacity, $R$ is thermal resistance, $c_p$ is specific heat capacity of the air, $g(t)$ is the number of people in a room and $p_g$ is internal heat gain per person. 

Similarly, we model the CO$_2$ dynamics described by~\cite{li2022tube, zhang2021novel}: 
\begin{equation}\label{eq:ODE_c}
\begin{aligned}
       m \frac{d C(t)}{d t} & =  (\alpha C_{\text{fresh}} + (1-\alpha) C(t) - C(t))u_{1,t} + \sigma g(t)
\end{aligned}
\end{equation}
where $m$ is the zone air mass, $\alpha$ is the supply air re-circulation rate, and $\sigma$ usually represents the average CO$_2$ generation rate per person.

The discrete version of the above ODE-based system can be written as:
\begin{equation}\label{eq:ODE}
\begin{aligned}
    & \begin{bmatrix}
        T(t+1) \\ C(t+1)
    \end{bmatrix} = \begin{bmatrix}
        T_e(t) \\ C(t)
    \end{bmatrix} + \\
    & \Delta t \cdot 
    \begin{bmatrix}
       \frac{1}{c} \left( \frac{T_{\text{ambient}}(t) - T_e(t)}{R} + c_p u_{1,t} (u_{2,t} - T_e(t)) + p_g g(t) \right) \\
       \frac{1}{m} \left( (\alpha (C_{\text{fresh}}  - C(t)))u_{1,t} + \sigma g(t)\right)
    \end{bmatrix}
\end{aligned}
\end{equation}

For the ODE-based control baseline, we employ ODE models~\eqref{eq:ODE} to model the average CO$_2$ level in a focus area $\Z_C$ and the average temperature in a room. The selection of $\Z_C$ targets areas typically exhibiting maximum CO$_2$ concentrations, aiming to model peak CO$_2$ levels and provide a fair comparison. 
Then we use the data-driven approach to learn $c, R, c_p, p_g, m, \alpha, \sigma$ and solve the formulated HVAC control problem:
\begin{subequations}
    \begin{align}
        \min_{u_0, u_1, \ldots, u_T} \quad &\sum_{t=0}^T
 Energy(t) \\
 & ~\eqref{eq:ODE} \tag{\text{ODE dynamics}} \\
 & C(t) \leq C_{\text{max}}, T_{\text{min}} \leq T_e(t) \leq T_{\text{max}} 
 \end{align}
\end{subequations}
where $Energy(t)$ is defined in~\eqref{eq:energy}. After solving the ODE-constrained optimization problem, the control actions are implemented on the building PDE environment.

\subsection*{Appendix B. Implementation details of the RL-based building control}\label{sec:app_b}
We formulate the HVAC control problem as a Markov Decision Process (MDP), which can be solved by RL. 
An MDP is composed of four elements: state ($s$), action ($a$), state transition probability ($p(s'|s, a)$), and reward ($r(s, a)$). The four elements are defined as follows: 
\begin{itemize}
    \item State: current temperature and CO$_2$ concentrations $s(t) = [T_e(z,t), C(z,t), \forall z \in \Z] \in \R^{n_x \times n_y \times 2}$.
    \item Action: $a(t) \in \R^2$ includes the supply airflow rate $a_{1,t}$ and supply air temperature $a_{2,t}$.
    \item State transition probability: Dynamics described by PDE in Section~\ref{sec:2}.
    \item Reward: the energy consumption cost plus the comfort and healthy violation cost which is defined as follows:
    \begin{equation}
    \begin{aligned}
        r(s(t), a(t)) &= - \|a(t)\|_1 \\
        & - \gamma_1 \max(\frac{1}{\Z}\int_z T_e(z,t)\text{d}z-T_{\text{max}}, 0) \\
        & - \gamma_1 \max(T_{\text{min}} - \frac{1}{\Z}\int_z T_e(z,t)\text{d}z, 0)\\
        & - \gamma_2 \max(\max_z C(z,t)-C_{\text{max}}, 0) \\
        & + \exp (1 - \max_z C(z,t)/400) \\
        & + \exp (-(21.5-\frac{1}{\Z} \int_z T_e(z,t) \text{d}z)^2)
    \end{aligned}
    \end{equation}
\end{itemize}
We apply the Proximal Policy Optimization (PPO)~\cite{schulman2017proximal} method to optimize control policy with $\gamma_1=5, \gamma_2=0.1$. We avoid directly using the negative of the loss~\eqref{eq:control_cost} as the reward to prevent NaN rewards. This issue arises because randomly explored actions in RL can cause temperature or CO$_2$ levels to exceed their limits, leading to negative inputs for the logarithm.
The PPO parameters are listed in Table~\ref{tab:ppo}. We normalize the action space and train the PPO model with stable-baselines3~\cite{stable-baselines3}.

\begin{table}[H]
\caption{Parameter for training the PPO model.}
\begin{tabular}{ll}\hline
PPO Parameter                                                                                    & Value  \\ \hline
Learning Rate                                                                                    & 0.0003 \\
Num Steps per Update                    & 2048   \\ 
Batch size                                                                                      & 64     \\ 
Num Epochs  per Surrogate Loss Update               & 10     \\ 
Discount Factor $\gamma$                                                                         & 0.99   \\
Clipping Parameter $\epsilon$                                                                    & 0.2    \\ 
Entropy Coefficient for Loss                                                                     & 0.0    \\ 
Value Function Coefficient for Loss               & 0.5    \\
Max Value for Gradient Clipping               & 0.5   \\\hline
\end{tabular}
\label{tab:ppo}
\end{table}